\documentclass[a4paper,12pt]{article}
\usepackage[cp1250]{inputenc}
\usepackage[T1]{fontenc}
\usepackage[slovak, english]{babel}
\usepackage{amssymb}
\input{epsf}
\begin{document}
\renewcommand{\abstractname}{\bf \small ABSTRACT}

\centerline{\large\bf  BEZIER CURVES INTERSECTION USING }

\centerline{\large\bf RELIEF PERSPECTIVE}
\bigskip
\centerline{\bf Radoslav Hlú\v{s}ek}
\bigskip
 \centerline{Department of Geometry}
 \centerline{Faculty of Mathematics and Physics}
\centerline{ Comenius University, Bratislava}
\centerline{Slovak Republik}
 \centerline{e-mail:
hlusek@fmph.uniba.sk}
\bigskip \bigskip
\begin{abstract}
{\footnotesize Presented paper describes the method for finding the intersection of class space
rational Bezier curves. The problem curve/curve intersection belongs among basic geometric
problems and the aim of this article is to describe the new technique to solve the problem using
relief perspective and Bezier clipping.}
\end{abstract}

\section*{\bf \large 1 Introduction}
\qquad Curve/curve intersection is one of the fundamental problems of computational geometry. At
the present time there exist several different approaches to this problem but the endeavor is to
avoid difficulties in calculations which are mainly results of polynomial representation rational
higher degree curves. It can be done in certain cases by using the relief perspective and Bezier
clipping.\\

Bezier clipping, in the context of plane curve intersection in this paper, see e.g.
\cite{nishita1} \cite{nishita2}, is an interactive method which takes advantages of the convex
hull property of Bezier curves. Regions of one curve which are guaranteed do not intersect a
second curve can be identified and subdivided away.\\

Relief perspective is a mapping of space into space in order to correspond conditions of the human
seeing. Images of all objects under the relief perspective are located in space between two
parallel planes.
\section*{\bf \large 2 Relief perspective}
\qquad The relief perspective is a special case of perspective collineation of the extended
Euclidean space $\bar {\mathbb{E}}_3$. We remind that a collineation is a bijective mapping
$\varphi:\bar{\mathbb{E}}_3\rightarrow \bar{\mathbb{E}}_{3}$ that preserves collinearity of
points. If there exists a plane $\omega$ such that $\varphi(X)=X$ $\forall X\in\omega$,\ $\varphi$
is a perspective collineation (for more details, see e.g. \cite{bus} \cite{cizm}). Then there
exists a point $O$ (called the centre of perspective collineation) such that $\varphi(O)=O$ and
$\varphi(\alpha)=\alpha$ $\forall \alpha;\ O\in\alpha$.\\

The relief perspective is a perspective co\-lli\-nea\-tion including some additional conditions:
\begin{itemize} \itemsep0pt \topsep0pt
  \item in order to produce correct images of 3D objects, it is important to respect necessary
  conditions of the human seeing. As in linear perspective, see e.g. \cite{cenek}, also in the
  relief perspective we
suppose, that objects are located inside viewing circular cone. The cone has a vertex in the eye
(the centre of projection) and the distance from the eye to the objects has to be at least 25 cm.
  \item the image plane $\omega$ (the set of invariant points ) does not
  contain the centre $O$ (i.e. $\varphi$ is a homology) and determine elements of the mapping
  $\varphi$ are: the centre $O$, the image plane $\sigma$ (the set of invariant points) and the
  vanishing plane $\omega^r$ (the image of the plane at infinity).
  \item no objects are ideal, i.e. objects and their images in the relief perspective have
  not ideal points. In notions of ${\mathbb{E}}_{3}$ the image plane has to be placed between
  the point $O$ and the plane $\omega^r$. The mapped object, that image we want to construct, is
  located in the semi-space, that it is opposite to the semi-space $\stackrel{\longmapsto}{\sigma O}$
  ("behind the image plane $\sigma$").
\end{itemize}
\quad Let $\varphi$ be a relief perspective of space $\bar {\mathbb{E}}_3$. Let denote $\bar
{\mathbb{E}}_3$ be a preimage space and $\bar {\mathbb{E}}_3^r$ be an image space ($\bar
{\mathbb{E}}_3$ with upper index "r") such that  $\varphi:\bar{\mathbb{E}}_3\rightarrow
\bar{\mathbb{E}}_{3}^r$. The image of the object ${\mathcal U}\in \bar {\mathbb{E}}_3$ under the
relief perspective $\varphi$ is called the relief of the object ${\mathcal U}$ and is denoted
analogically ${\mathcal U}^r$. The relief perspective is given by the centre $O$, the image plane
$\sigma$ and ordered pair of different points $A,A^r= \varphi(A)\ (A,\varphi(A)\neq O)$ such that
$A,A^r \not\in\sigma$ and $O,A,A^r$ are collinear.\\

Note therefore all mapped objects ${\mathcal U}$ and their images ${\mathcal U}^r$ under the
relief perspective have no ideal points, we can study them in ${\mathbb{E}}_3$ or
$\bar{\mathbb{E}}_3$ using affine and projective methods as well. It is also clear that the
restriction $\varphi/{\mathcal U}$ of the relief perspective $\varphi$, which is a bijective
mapping, is again a bijection.
\section*{\bf \large 3 Representation of the re\-lief pers\-pec\-ti\-ve in an analytical form}
\qquad Suppose $X$ and $X'$ are two different points of
${\mathbb{E}}_{3}\subseteq\bar{\mathbb{E}}_3$ with coordinates $[x,y,z]$ and $[x',y',z']$
respectively. If $\varphi$ is a collineation of the space $\bar {\mathbb{E}}_3$ we use the
following equations:
\begin{eqnarray}
\label{collineation}
  x'=\frac{a_{11}x+a_{12}y+a_{13}z+a_{14}}{a_{41}x+a_{42}y+a_{43}z+a_{44}}\nonumber  \\
  y'=\frac{a_{21}x+a_{22}y+a_{23}z+a_{24}}{a_{41}x+a_{42}y+a_{43}z+a_{44}} \\
  z'=\frac{a_{31}x+a_{32}y+a_{33}z+a_{34}}{a_{41}x+a_{42}y+a_{43}z+a_{44}}, \nonumber
\end{eqnarray}
where $a_{i,j}$ $(i,j\in\{1..4\})$ are real numbers and $ det(a_{ij})\neq0$. In equations
(\ref{collineation}) are used non-homogenic coordinates, which correspond to a restriction
$\bar\varphi=\varphi/\{\bar{\mathbb{E}}_3-\alpha\}$, where $\alpha$ is a plane
$a_{41}x_1+a_{42}x_2+a_{43}x_3+a_{44}x_0=0$.\\

In order to obtain simple mapping equations for relief perspective from (\ref{collineation}) we
assume, that the centre $O$ is a point  with coordinates $[0,0,0]$ and $\sigma$ is a plane
$z-1=0$. In this special case we obtain the convenient representation of the relief perspective in
the following form
\begin{eqnarray}
\label{rel persp rovnice} x'=\frac{(1+k)x}{z+k}\nonumber\\ y'=\frac{(1+k)y}{z+k}\\
z'=\frac{(1+k)z}{z+k}\nonumber
\end{eqnarray}
where the span $k\in{\mathbb R}^+-\{1\}$ 
(the geometric representation of the parameter $k$ will be explained in the next lines).
\section*{\bf \large 4 Some properties of the relief perspective}
\begin{itemize}
\item the vanishing plane $\omega^r\equiv 1+k$ and the neutral plane $\nu\equiv -k$
(the preimage of the plane at infinity) are parallel and for distances $ O,\nu $ and
$\omega^r,\sigma $ we have $$\mid O\nu\mid=\mid\omega^r\sigma\mid=k$$ where the parameter $k$
represents the span
\item if a plane $\alpha\equiv z-c\ (c\ \neq 0)$ is parallel to the image plane
$(\alpha\neq\sigma,\nu)$, then the relief of the plane $\alpha$ is the plane
$\alpha^r\equiv\frac{(1+k)c}{c+k}$ and these planes are parallel (it is obvious that
$\alpha\parallel\alpha^r\parallel\sigma$)
\item the relief of a point placed in the semi-space, that is opposite to the semi-space
$\stackrel{\longmapsto}{\sigma O}$, is a point in a part of space with boundary planes $z=1$ and
$z=1+k$ (in an intersection of spaces $z>1$ and $z<1+k$) (Fig. \ref{pohlad})
\begin{figure}[htbp]
\begin{center}
\leavevmode \epsfxsize =8cm
\epsfbox{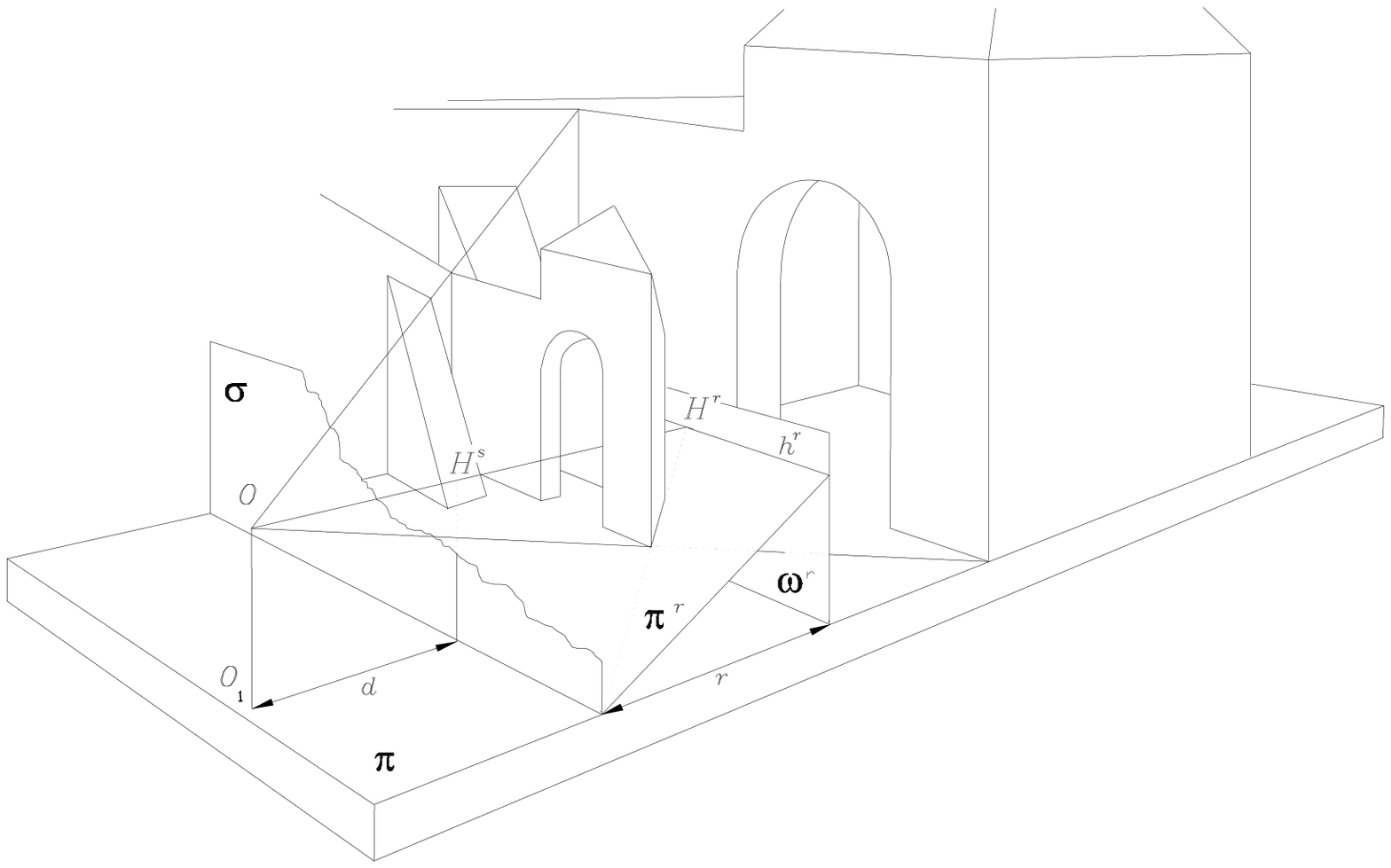}
\end{center}
\renewcommand{\caption}[1]{\refstepcounter{figure} \centering Figure \thefigure}
\caption{} \label{pohlad}
\end{figure}
\item the orthographic projection of the object ${\mathcal U}^r$ (the relief of the
object $\mathcal U$) onto the image plane $\sigma$ is the central projection of the object
$\mathcal U$ from the point $S$, where $S\in l\cap\nu\quad (l;O\in l\wedge l\perp\sigma)$
\end{itemize}
\section*{\bf \large 5 Rational Bezier curves and relief perspective}
\qquad The relief perspective is a mapping of the extended Euclidean space $\bar {\mathbb{E}}_3$.
Let $V_i,\ i\in\{0,1,\ldots,n\}$ be given points, $[x_i,y_i]$ be their coordinates
 of the space $\bar {\mathbb{E}}_2$ and positive real numbers $w_i,\
i\in\{0,1,\ldots,n\}$ be their weights. By these elements and Bernstein polynomials $B_i^n(t)$,
see e.g. \cite{farin}, is defined a n-th planar rational Bezier curve
\begin{eqnarray}
\label{2D rac BK}
{\bf {P}}(t)=
[\frac{\sum\limits_{i=0}^nx_iw_iB_i^n(t)}{\sum\limits_{i=0}^nw_iB_i^n(t)},\,
\frac{\sum\limits_{i=0}^ny_iw_iB_i^n(t)}{\sum\limits_{i=0}^nw_iB_i^n(t)}],\quad
 t\in \langle0,1\rangle
\end{eqnarray}
and also the space nonrational Bezier curve
\begin{eqnarray}
\label{nerac priestor BK}
{\bf\bar{P}}(t)=[\sum_{i=0}^nx_iw_iB_i^n(t),\,\sum_{i=0}^ny_iw_iB_i^n(t),\,\sum_{i=0}^nw_iB_i^n(t)],\qquad
t\in \langle0,1\rangle
\end{eqnarray}
with control points
\begin{equation}
\label{3d body z 2d a vah} V_i=[x_iw_i,\,y_iw_i,\,w_i].
\end{equation}
\quad Let  $\varphi:\bar{\mathbb{E}}_3\rightarrow \bar{\mathbb{E}}_{3}^r$ be the relief
perspective of the space  $\bar{\mathbb{E}}_{3}$ and $X^r\/[x,y,z]$ $(X^r\not\in\sigma)$ be the
relief of the point $X[a,b,c]\in \bar{\mathbb{E}}_{3}$. From the equations (\ref{rel persp
rovnice}) we get
\begin{equation}
x=\frac{(1+k)a}{c+k},\  y=\frac{(1+k)b}{c+k},\ z=\frac{(1+k)c}{c+k}
\end{equation}
and according to the relation $z=\frac{(1+k)c}{c+k}$ we obtain
\begin{eqnarray}
c+k=\frac{k(1+k)}{1+k-z}\nonumber
\end{eqnarray}
From the above results coordinates $[a,b,c]$ of the point $X$, that is the preimage of the relief
$X^r[x,y,z]$, are computed according to $a=\frac{kx}{1+k-z}$, $b=\frac{ky}{1+k-z}$,
$c=\frac{kz}{1+k-z}$.\\ \rule[1mm]{0pt}{1mm}\\ {\bf Lemma} {\em Let
$\varphi:\bar{\mathbb{E}}_3\rightarrow\bar{\mathbb{E}}_{3}^r$ be relief perspective of the space
$\bar{\mathbb{E}}_{3}$. The point $[x,y,z]$ is the relief of the point
$[\frac{kx}{1+k-z}$,$\frac{ky}{1+k-z}$,$\frac{kz}{1+k-z}]$.}\\ \rule[1mm]{0pt}{1mm}\\

By given planar points $V_i$ and 
their weights $w_i$, where $1+k>w_i \ \forall i$, we can define in $\bar{\mathbb{E}}_3$ points
\begin{equation}
\label{vrcholy nerac BK}
W_i=[\frac{kx_iw_i}{1+k-w_i},\,\frac{ky_iw_i}{1+k-w_i},\,\frac{kw_i}{1+k-w_i}]\qquad
i\in\{0,1,\ldots,n\}
\end{equation}
With respect to the lemma above it is obvious, that the reliefs of the points $W_i$ are
represented by the points (\ref{3d body z 2d a vah}). Using the points $V_i$, the space
nonrational Bezier curve ${\bf \bar{P}}(t)$ of the form (\ref{nerac priestor BK}) is defined and
we notice, that ${\bf \bar{P}}(t)$ under mapping $\varphi$ is an image of the curve
\begin{equation}
\label{rac priestor k BK}
{\bf Q}(t)=
[\frac{\sum\limits_{i=0}^nkx_iw_iB_i^n(t)}{\sum\limits_{i=0}^n(1+k-w_i)B_i^n(t)},\
\frac{\sum\limits_{i=0}^nky_iw_iB_i^n(t)}{\sum\limits_{i=0}^n(1+k-w_i)B_i^n(t)},\,
\frac{\sum\limits_{i=0}^nkw_iB_i^n(t)}{\sum\limits_{i=0}^n(1+k-w_i)B_i^n(t)}]
\end{equation}
This curve {\bf Q}(t) is the space rational Bezier curve defined by control points of the form
(\ref{vrcholy nerac BK}) and their weights $\Omega_i$, that are computed according to
$\Omega_i=1+k-w_i$ because of
$$\frac{kx_iw_i}{1+k-w_i}\Omega_i=kx_iw_i,\quad\frac{ky_iw_i}{1+k-w_i}\Omega_i=ky_iw_i,\quad
\frac{kw_i}{1+k-w_i}\Omega_i=kw_i$$
\quad Applying all these results we obtain the following
theorem\\ \rule[1mm]{0pt}{1mm}\\ {\bf Theorem} {\em The curve ${\bf \bar{P}}$(t) is a relief of
the curve {\bf Q}(t).}\\ \rule[1mm]{0pt}{1mm}\\

\qquad In order to get more information about Bezier curves and the relief perspective, let us
assume having a central projection with the centre $O[0,0,0]$ and the plane $z-1=0$. In this case
the planar rational curve ${\bf {P}}(t)$ of the form (\ref{2D rac BK}) is the image of the space
rational Bezier curve ${\bf Q}(t)$ defined by (\ref{rac priestor k BK}). \\

Let the space nonrational Bezier curve ${\bf \bar{P}}(t)$ expressed by (\ref{nerac priestor BK})
be given. What is the relief of this curve? We know $z+k=\sum\limits_{i=0}^n(w_i+k)B_i^n(t)$ and
now the relief ${\bf \bar{P}^r}(t)$ of the given curve ${\bf P^r}(t)$ can be written as
\begin{equation}
\label{relief priestor nerac BK}  {\bf \bar{P}^r}(t)=
[\frac{\sum\limits_{i=0}^n(1+k)x_iw_iB_i^n(t)}{\sum\limits_{i=0}^n(w_i+k)B_i^n(t)},\,\nonumber\\
\frac{\sum\limits_{i=0}^n(1+k)y_iw_iB_i^n(t)}{\sum\limits_{i=0}^n(w_i+k)B_i^n(t)},\,
\frac{\sum\limits_{i=0}^n(1+k)w_iB_i^n(t)}{\sum\limits_{i=0}^n(w_i+k)B_i^n(t)}]
\end{equation}
$t\in\langle0,1\rangle,$\qquad where control vertices are points expressed as follow $$
\bar{V}_i^r=[\frac{(1+k)x_iw_i}{w_i+k},\,\frac{(1+k)y_iw_i}{w_i+k},\,\frac{(1+k)w_i}{w_i+k}]$$ and
weights $\bar{\Omega}_i$ are computed by $\bar{\Omega}_i=w_i+k$.

\section*{\bf \large 6 Intersection of space rational Bezier curves}
\qquad Let ${\bf P}(t)$, ${\bf Q}(u)$ be space rational Bezier curves of the form (\ref{rac
priestor k BK}). The aim is to find their intersection.\\

\begin{figure}[htbp]
\begin{center}
\leavevmode \epsfxsize = 8.5cm
\epsfbox{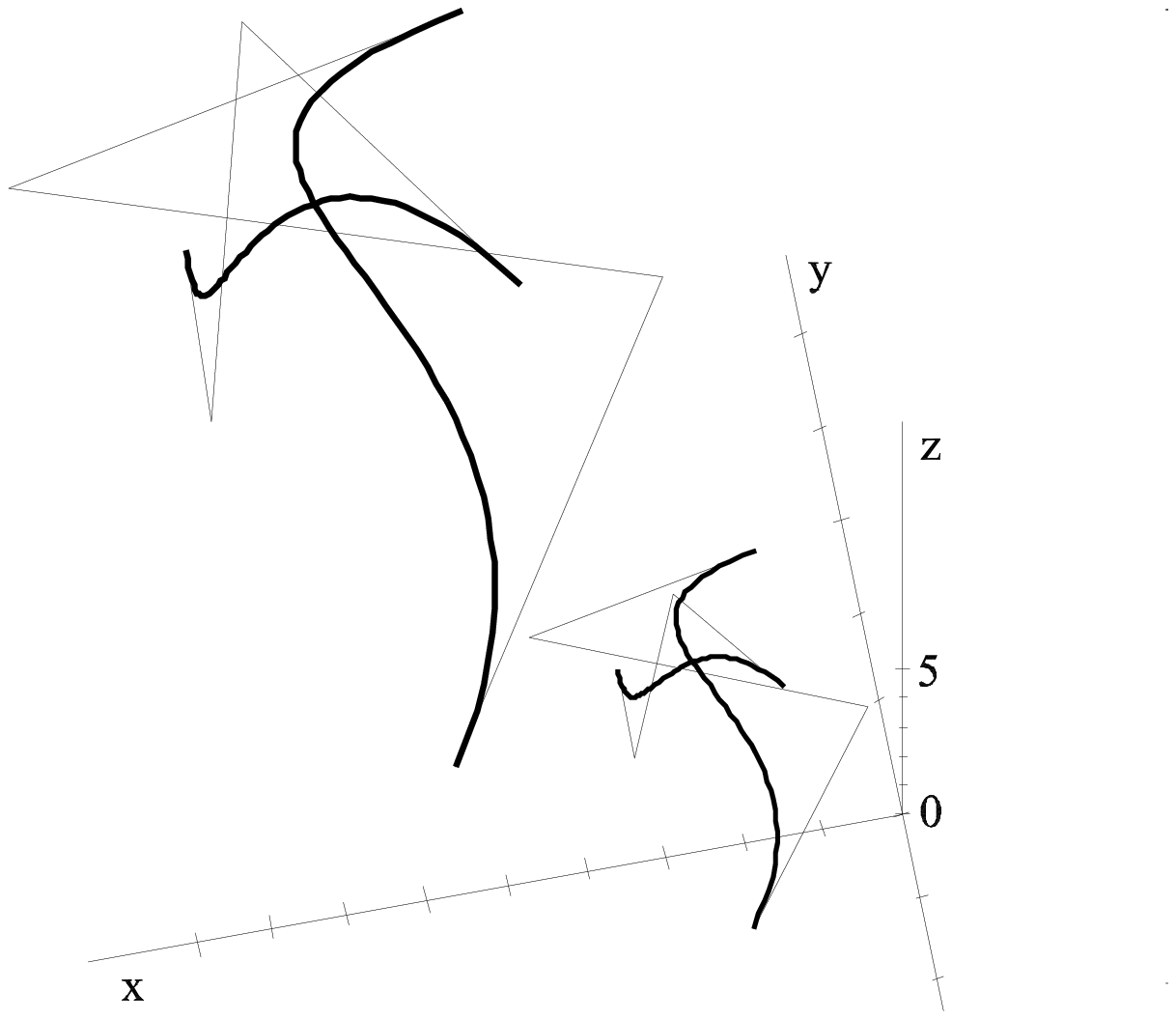}
\end{center}
\renewcommand{\caption}[1]{\refstepcounter{figure} \centering Figure \thefigure}
\caption{} \label{inters}
\end{figure}

The solution to this problem can be found using already known facts. According to the theorem,
which was formed in the previous section, reliefs of space rational Bezier curves of the form
(\ref{rac priestor k BK}) are space nonrational Bezier curves of the form (\ref{nerac priestor
BK}) (in Fig. \ref{inters} Bezier curves of the form (\ref{rac priestor k BK}) and their reliefs
for $k=2$ are shown).
 The central projections of these
nonrational curves from the point $[0,0,0]$ onto the plane $z-1=0$ are planar rational Bezier
curves defined by (\ref{2D rac BK}). Their intersection is possible to find specifically by using
the method of Bezier clipping. This method specifies values of the parameter
$t\in\langle0,1\rangle$ of the curve ${\bf P^s}(t)$, respectively the parameter
$u\in\langle0,1\rangle$ of the curve ${\bf Q^s}(u)$, that correspond to the common point ${\bf
R^s}$ (the upper indexes "r" and "s" denote the reliefs and central projections of curves or
points). This point is the intersection of both curves . The preimage of the point ${\bf R^s}$ in
the central projection is the space point, which is the relief of the intersection of the space
rational Bezier curves ${\bf P}(t)$ and ${\bf Q}(u)$.\\

The following scheme shows a whole proces of finding the intersection of the ${\bf P}(t)$ and
${\bf Q}(u)$ curves: $$\left.
\begin{array}{llr}{\bf P}(t)\stackrel{\varphi}{\longrightarrow}{\bf
P^r}(t)\stackrel{\psi}{\longrightarrow}{\bf P^s}(t) \\ {\bf
Q}(u)\stackrel{\varphi}{\longrightarrow}{\bf Q^r}(u)\stackrel{\psi}{\longrightarrow}{\bf
Q^s}(u)\end{array}\right\} \stackrel{\mbox{\tiny B. clipping}}{\longrightarrow}{\bf R^s}$$ $${\bf
R^s}={\bf {P^s}}(t)\cap{\bf {Q^s}}(u) \stackrel{\psi^{-1}}{\longrightarrow}{\bf
R^r}\stackrel{\varphi^{-1}} {\longrightarrow}{\bf R}\quad$$ $${\bf R}={\bf {P}}(t)\cap{\bf
{Q}}(u)$$ $\varphi$ -- relief perspektive
$\varphi:\bar{\mathbb{E}}^3\rightarrow\bar{\mathbb{E}}^{3}$\\ $\psi$ -- central projection
$\psi:\bar{\mathbb{E}}^3\rightarrow\bar{\mathbb{E}}^{2}$.\\

The central projection $\psi$ is not a bijective mapping and in case that the intersection of
curves ${\bf {P^s}}(t)$ and ${\bf {Q^s}}(u)$ exists (in opposite case the given curves ${\bf
P}(t)$ and ${\bf Q}(u)$ certainly do not intersect) the intersection of curves ${\bf P}(t)$ and
${\bf Q}(u)$ does not have to exist. This situation occurs when preimages of point ${\bf R^s}$ on
the curves ${\bf {P^r}}(t)$ and ${\bf {Q^r}}(u)$ in the central projection are different.

\section *{\bf \large 7 Conclusions and future work}
\qquad The relations between Bezier curves and the relief perspective have been described. The
necessary and sufficient conditions for expressing the space nonrational Bezier curve as the
relief of the space rational Bezier curve have been formed. We have shown that to the planar
rational curve of the form (\ref{2D rac BK}), defined by planar points $V_i\in\bar{\mathbb{E}}_2$
and their weights $w_i$, it is possible to assign a class of the space curves by the relief
perspective. One of them is nonrational curve defined by (\ref{nerac priestor BK}) and two are
rational curves defined by (\ref{rac priestor k BK}) and (\ref{relief priestor nerac BK}).\\

The described method can be used as a direction for application how to find the intersection of
the space rational curves of the form (\ref{rac priestor k BK}). \quad It is possible to express
every polynomial curve in Bezier's representation and due to this representation the method can be
applied to all polynomial curves after modifications (e.g. spline curves which are considered as
curves consist of Bezier segments) for solution to the curve/curve or curve/line intersection
problems.\\

Despite the author's attempt he did not succeed in finding similar comparable published methods on
website. In addition to this he is not able to compare his results with any other. Obviously
author's knowledge is limited and he would appreciate to get information about any other similar
method.\\

In the future work we want to extend possibilities of the relief perspective in geometric
modeling. Our aim is to find an answer whether any polynomial 3D curve can be converted to the
space curve of the form (\ref{rac priestor k BK}).
\renewcommand{\refname}{\bf References \footnote{some of the publications have not equivalents
in English}}

\end{document}